\documentclass{article}

\usepackage{arxiv}

\usepackage[utf8]{inputenc} 
\usepackage[T1]{fontenc}    
\usepackage{hyperref}       
\usepackage{url}            
\usepackage{booktabs}       
\usepackage{amsfonts}       
\usepackage{nicefrac}       
\usepackage{microtype}      
\usepackage{lipsum}
\usepackage{graphicx}
\usepackage{amsmath}
\graphicspath{ {./images/} }

\title{Experimental Demonstration of Millimeter-Wave Radio-over-Fiber System with Convolutional Neural Network (CNN) and Binary Convolutional Neural Network (BCNN)}

\author{
 Jeonghun Lee \\
  School of Engineering\\
  RMIT University\\
  Melbourne, VIC, AU, 3000 \\
  \texttt{s3638815@student.rmit.edu.au} \\
   \And
 Jiayuan He \\
  School of Information and Computer Science\\
  University of Melbourne\\
  Melbourne, VIC 3010 \\
  \texttt{hjhe@student.unimelb.edu.au} \\
  \And
 Yitong Wang \\
  School of Engineering\\
  RMIT University\\
  Melbourne, VIC, AU, 3000 \\
  \texttt{s3587959@student.rmit.edu.au} \\
  \And
 Chengwei Fang \\
  School of Engineering\\
  RMIT University\\
  Melbourne, VIC, AU, 3000 \\
  \texttt{s3643273@student.rmit.edu.au} \\
  \And
 Ke Wang \\
  School of Engineering\\
  RMIT University\\
  Melbourne, VIC, AU, 3000 \\
  \texttt{ke.wang@rmit.edu.au} \\
}

\begin{document}
\maketitle
\begin{abstract}
The millimeter-wave (mm-wave) radio-over-fiber (RoF) systems have been widely studied as promising solutions to deliver high-speed wireless signals to end users, and neural networks have been studied to solve various linear and nonlinear impairments. However, high computation cost and large amounts of training data are required to effectively improve the system performance. In this paper, we propose and demonstrate highly computation efficient convolutional neural network (CNN) and binary convolutional neural network (BCNN) based decision schemes to solve these limitations. The proposed CNN and BCNN based decision schemes are demonstrated in a 5 Gbps 60 GHz RoF system for up to 20 km fiber distance. Compared with previously demonstrated neural networks, results show that the bit error rate (BER) performance and the computation intensive training process are improved. The number of training iterations needed is reduced by about 50 \% and the amount of required training data is reduced by over 30 \%. In addition, only one training is required for the entire measured received optical power range over 3.5 dB in the proposed CNN and BCNN schemes, to further reduce the computation cost of implementing neural networks decision schemes in mm-wave RoF systems.
\end{abstract}


\section{Introduction}
Radio-over-fiber (RoF) systems have been widely studied to support high speed wireless communications by leveraging the advantages of optical fibers, such as the low transmission loss and the broad bandwidth. However, mm-wave RoF systems are typically impaired by a number of linear and nonlinear effects, induced during signal modulation, amplification, transmission and detection \cite{Impair1}. To overcome the limitations in mm-wave RoF systems, various technologies have been proposed and investigated. To reduce the impact of fiber dispersion and to extend the fiber reach, the double sideband with carrier suppression (DSB-CS) scheme has been proposed and studied \cite{Fiber_DSBCS}. In addition, signal processing techniques have been proposed at the receiver side to improve the system performance \cite{mmwave_dsp1}, where digital signal processing has been widely studied to mainly suppress various linear impairments and analog circuits have been studied to alleviate nonlinear effects \cite{DSP_impair}\cite{analog_tech}.  However, in traditional signal processing schemes, each processing step typically only targets at one or few impairments \cite{DSP_impair} and solving interactions amongst impairments is challenging. Furthermore, most signal processing methods target at linear impacts, whilst nonlinear effects are more challenging to be suppressed.

\begin{figure}[b]
\centering\includegraphics[width=84mm]{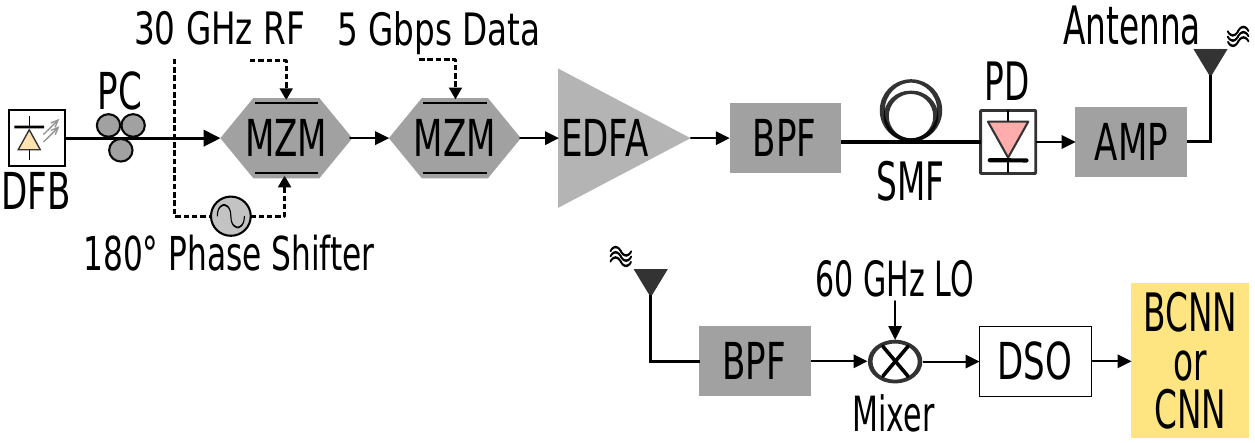}
\caption{Architecture of the millimeter-wave RoF system with proposed CNN and BCNN}
\label{fig:figure1}
\end{figure}

On the other hand, neural networks have been proposed and studied in optical communication systems recently, including in mm-wave RoF systems \cite{ML_ex_mmWaveRoF}. Neural networks are widely considered as equalizers and classifiers in optical communication systems \cite{ML_ex2_RNN}. Compared with traditional signal processing methods, neural networks are capable of compensating nonlinear effects and processing various impairments simultaneously. Due to these unique advantages, neural networks have achieved better system performance \cite{ML_ex4}. It has been demonstrated that as a classifier, the neural network is capable of mitigating both linear and nonlinear impairments in a mm-wave RoF system with DSB-CS \cite{ML_ex_mmWaveRoF}.

However, most previously demonstrated neural network schemes are based on the fully-connected architecture \cite{ML_ex4}, which has a large number of learnable parameters and requires high computation cost, especially during training processes. Therefore, their practical applications are significantly limited, especially using low-profile platforms.

In this paper, we propose and demonstrate the convolutional neural network (CNN) and binary convolutional neural network (BCNN) based decision schemes in mm-wave RoF systems to significantly reduce the computation requirement by adopting 1-D convolution and maxpooling computations. Experimental results show that in a 5 Gbps 60 GHz RoF system with DSB-CS modulation, the CNN and BCNN based decision schemes can significantly improve the most computation-intensive training process. Compared with previously demonstrated neural networks, about 50 \% reduced training iterations and better BER performance are achieved simultaneously. The required training data is also reduced by over 30 \% for both CNN and BCNN. Therefore, the computation cost of using neural network principles in mm-wave RoF systems is significantly reduced. In addition, only one training is required in the proposed CNN and BCNN schemes over the entire over 3.5 dB measured received optical power range, and hence, the computation requirement of both neural networks are further significantly reduced.

\begin{figure}[!t]
\centering\includegraphics[width=84mm, height=70mm]{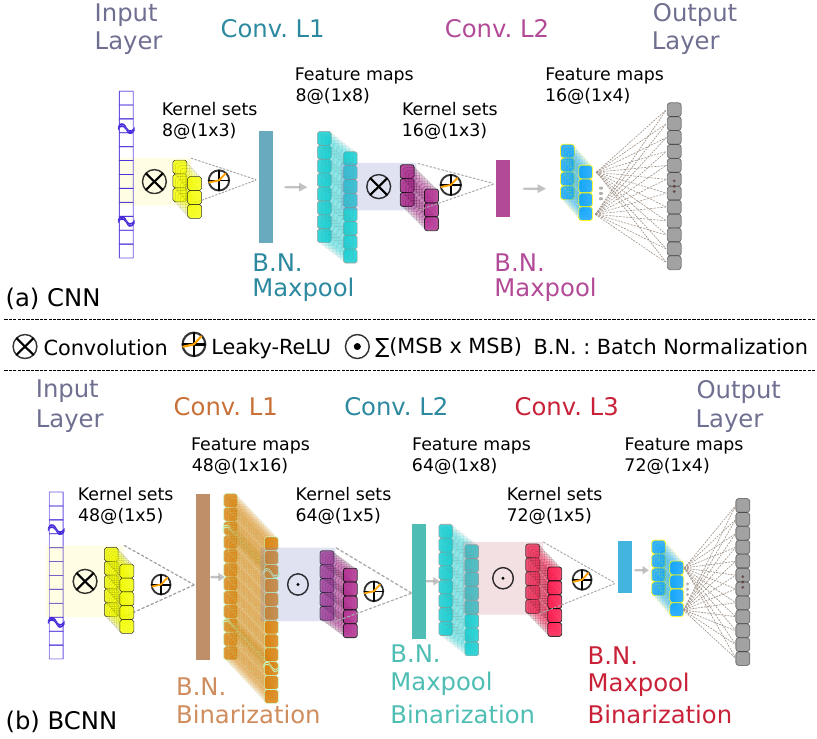}
\caption{Architecture of the proposed neural networks for the RoF systems. (a) CNN; and (b) BCNN}
\label{fig:CNN_BCNN_Archit}
\end{figure}

\section{Architecture of Mm-Wave RoF System with CNN and BCNN}
The 60 GHz mm-wave RoF system considered is shown in Fig.\ref{fig:figure1}. The DSB-CS scheme is used, since it suffers from less fiber dispersion compared with other RoF schemes. A DFB laser at the 1550 nm band is used as the light source, and a dual-driven Mach-Zehnder modulator (MZM) biased at $\ V_{\pi}\ $ and driven by two complementary 30 GHz RF signal is used to realize the optical carrier suppression. The data to be transmitted is then modulated using a second MZM. The signal is amplified by an Erbium-doped Fiber Amplifier (EDFA) and filtered by an optical bandpass filter (BPF) before transmission via standard single mode fiber (SMF). After fiber transmission, the optical signal is detected by a PIN photodetector (PD). After bandpass filtering, RF amplification (AMP), wireless transmission and down-conversion, the received signal is then fed into the proposed CNN- and BCNN-based decision schemes.

The architecture of proposed CNN decision scheme is shown in Fig.\ref{fig:CNN_BCNN_Archit}(a), which consists of the input layer, the convolutional layers, and the output layer. Here we consider received symbols as 1-dimensional (1-D) vectors in the input layer, which allows the use of computation efficient 1-D convolution operation to capture various system impairments information carried by received symbols in the subsequent convolutional layers. Each convolutional layer also carries out the nonlinear activation function and maxpooling. The 1-D convolution computation with kernel sets in the CNN decision scheme can be expressed as
\begin{equation}
Conv = \displaystyle \sum_{1}^{N}\sum_{1}^{R}\sum_{1}^{F}X \otimes \theta
\end{equation}
where X is the input data of each convolutional layer, $\theta$ is the kernel set, Conv is the outcome of the convolutional computation, N is the number of kernel sets, R is the data size , and F is the kernel size. By conducting the convolution computation, signal characteristics carried by multiple symbols can be effectively learnt and impairments amongst symbols can be suppressed correspondingly, such as the inter-symbol interference (ISI). In the convolutional layer, batch normalization is used to improve the accuracy and avoid overfitting issues, and the Leaky-ReLU is used as the activation function, which can be expressed as

\begin{equation}
\label{eq_leaky_relu}
    f(x) = \begin{cases}
               0.2\cdot x & x < 0\\
               x          & x \geq 0\\
           \end{cases}
\end{equation}
where \textit{x} is the result of convolution computations. We have implemented different activation functions in the CNN decision scheme, and we found that amongst Sigmoid, Tanh, ReLU and Leaky-ReLU functions, the Leaky-ReLU provides the least performance fluctuation, solves the gradient vanishing problem \cite{grad_vanishing}, and achieves the best BER performance. Therefore, we select the Leaky-ReLU function in convolutional layers. The optimized kernel parameters are determined by minimizing the loss function during the learning process. Through this way, the nonlinear impairments in the RoF system are suppressed. In the output layer of the CNN scheme, matrix-multiplication and accumulation operations are carried out on the output of convolutional layers to generate the final symbol decision.

In addition to the CNN decision scheme, we also propose a BCNN decision scheme to further reduce the computation cost. Here, we replace the 32-bit floating point number computations with the most-significant-bit (MSB) multiplications and additions, which can be expressed as
\begin{equation}
Binary-Conv
= \displaystyle \sum_{1}^{N}\sum_{1}^{R}\sum_{1}^{F}MSB\left(X\right) \times MSB\left(\theta\right)
\label{eq_bcnn}
\end{equation}

\begin{figure}[!t]
\centering\includegraphics[width=60mm, height=45mm]{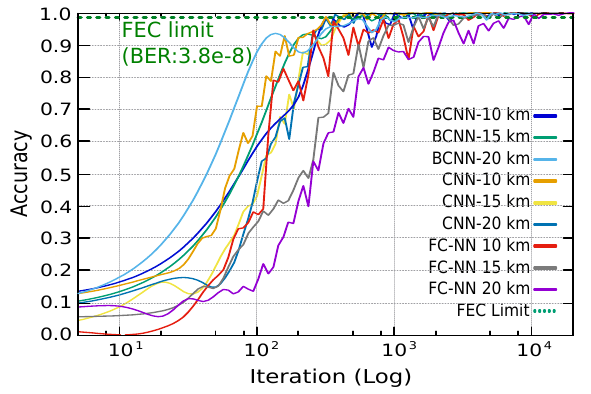}
\caption{CNN, BCNN, and FC-NN accuracy versus iteration results.}
\label{fig:training_result}
\end{figure}
The architecture of proposed BCNN based decision scheme is illustrated in Fig.\ref{fig:CNN_BCNN_Archit}(b). It is similar to the CNN based scheme except that it is implemented with binary convolution computation and the output of batch normalization is binarized. Leaky-ReLU activation function and maxpooling outputs are also converted into binary values. By binarizing parameters and outputs of each convolutional layer, as expressed by Eq. (\ref{eq_bcnn}), matrix multiplications and additions in both convolution and output layers are only computed with the sign bits of input data and kernel parameters. Thus, the computation cost is further significantly reduced and the BCNN-based decision scheme is more hardware friendly, especially for portable platforms with limited computation resources and capabilities.

\section{EXPERIMENTS AND RESULTS}
The proposed DSB-CS mm-wave RoF systems with CNN and BCNN decision schemes were experimentally investigated. The bit rate was 5 Gbps with the 2-PAM modulation format, the optical carrier wavelength was 1550 nm, and the optical power after amplification launched into the optical fiber was 3 dBm. Fiber transmission lengths of 10 km, 15 km and 20 km were measured in the experiment. After transmission via the optical fiber, the signal was detected by a 60 GHz PIN photodetector. Due to device limitations, the wireless transmission was not included, and the converted electrical signal was down-converted directly by a RF mixer. The received data was sampled and stored in a digital sampling oscilloscope (DSO), where the up-sampling factor was 4. The CNN- and BCNN-based decision schemes were then used to process the received signal.

One million symbols were transmitted and received, and they were divided into the training dataset and the testing dataset. Each symbol sequence consisted of 4 symbols (i.e., 16 sampled data points), and they were fed to the input layer. In the CNN-based decision scheme, 2 convolutional layers were used, and the number of feature sets was 8 and 16, respectively. The kernel size was fixed at 1 $\times$ 3. In the BCNN-based decision scheme, 3 convolutional layers were used, and they had feature sets of 48, 64, and 72, respectively. A larger kernel size of 1 $\times$ 5 was used to improve the accuracy. Although the BCNN had more layers, larger number of feature sets and larger kernel size, it still had lower computation cost compared to CNN due to the use of the MSB multiplication and addition operations. In addition, as a comparison, the previously demonstrated FC-NN was also implemented \cite{ML_ex4}, which had the input layer, 4 hidden layers, and the output layer. The number of nodes were 56, 60, 64, and 52, respectively.

\begin{figure}[!t]
\centering\includegraphics[width=60mm, height=135mm]{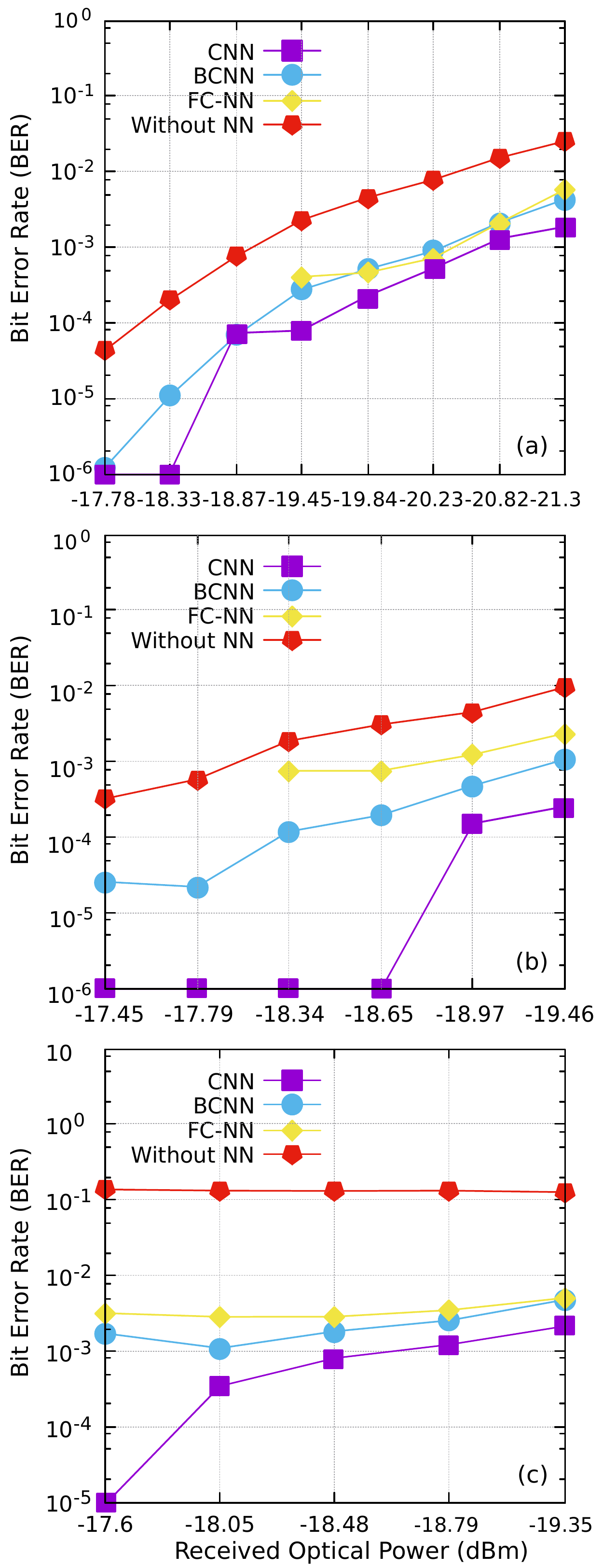}
\caption{CNN and BCNN processing results. BER over various fiber transmission lengths - (a) 10 km, (b) 15 km, (c) 20 km}
\label{fig:nn_result_distance}
\end{figure}
The proposed CNN, BCNN and the reference FC-NN based decision schemes were coded with the Tensorflow Framework. The loss function used during the training procedure was the softmax-cross-entropy, which can be expressed as.
\begin{equation}
Softmax-cross-entropy
= \displaystyle \frac{1}{N}\cdot(-\sum_{i}(L \cdot \log{Y}))
\label{softmax}
\end{equation}
where N is the number of output categories, i is the index of the output label, L is the one-hot encoded output label, and Y is the probability of output label L. The batch size was experimentally optimized and selected at 1024. The Adams optimization algorithm was used for training \cite{Adam_Opt} and the learning rate was 0.0005.

The training process is shown in Fig.\ref{fig:training_result}, which shows the accuracy with respect to iteration. It can be seen that both CNN and BCNN could achieve high accuracy of 98.5 \% (corresponding to the BER of 3.8 $\times$ 10 $^{-3}$ FEC limits) by about 50 \% fewer training iterations for all fiber transmission distances than FC-NN.

In the experiment, we characterized the BER of the mm-wave RoF system with respect to the received optical power. The measured BER performance with the proposed CNN- and BCNN-based decision schemes is shown in Fig.\ref{fig:nn_result_distance}.
It can be seen from the results that for the fiber transmission distance of 10 km, the proposed CNN-based decision schemes achieve improved BER performance  compared with previous FC-NN based scheme, while the BCNN-based scheme shows similar performance. Compared with the system without neural network, the BER performance is improved by over one order of magnitude, and the receiver sensitivity (defined at the BER of 3.8 $\times$ 10$^{-3}$) is improved by over 1.5 dB. For the fiber transmission distance of 15 km, it is clear from Fig.\ref{fig:nn_result_distance}(b) that both CNN and BCNN can achieve better BER performance than the FC-NN within the FEC limit over the entire measured received optical power range, and the BER improvement compared with the system without neural networks becomes larger than the 10 km fiber transmission scenario. Specifically, the CNN-based scheme achieves more than 1 order magnitude performance improvement than the BCNN scheme, which is mainly due to the loss of accuracy in BCNN during the binarization process. When the fiber transmission distance further increases to 20 km, the system without neural networks cannot achieve a BER within the FEC limit any more, whilst the systems with CNN and BCNN can still realize BER better than the FEC limit. Therefore, the CNN based scheme is better situated for longer transmission distances, even though the structure is less deep.

\begin{figure}[!t]
\centering\includegraphics[width=60mm, height=43mm]{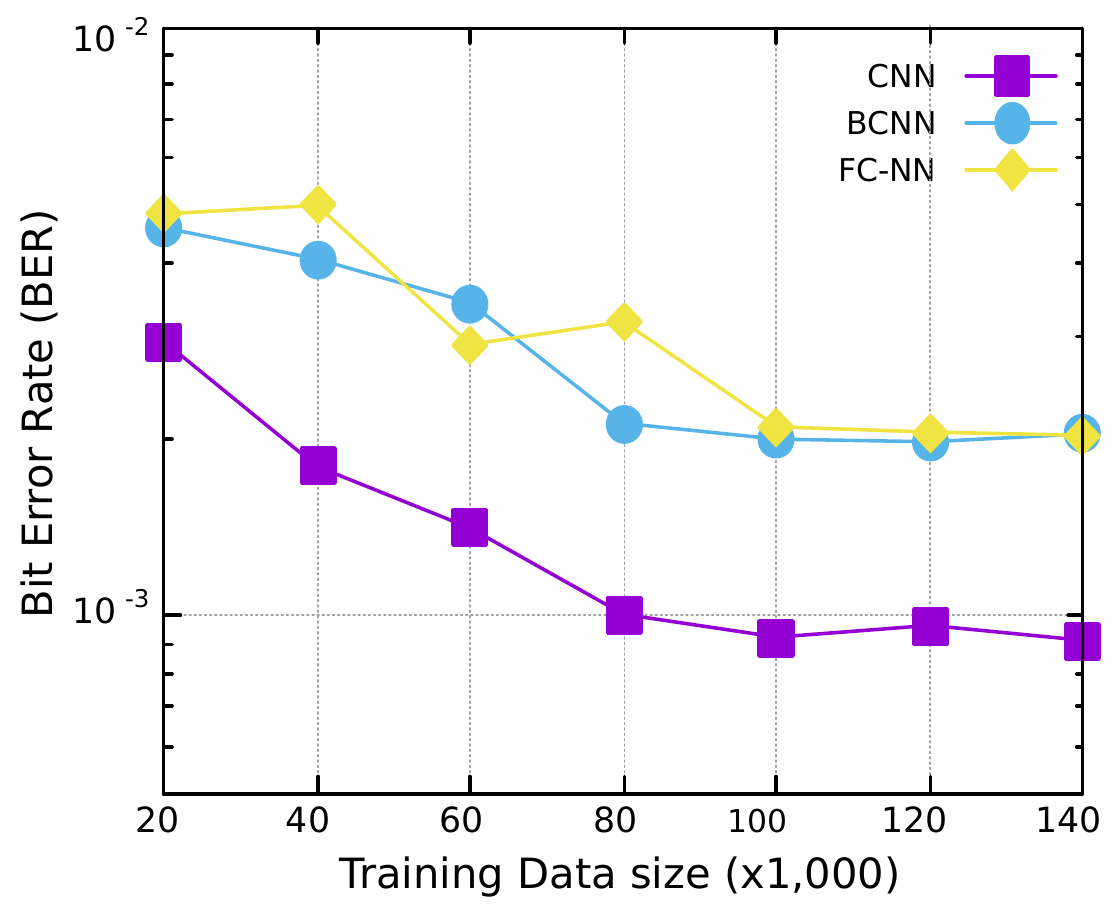}
\caption{BER versus training data sizes simulation result - fiber distance of 10km at the receiver power of -17.78 dBm}
\label{fig:ber_vs_datasize}
\end{figure}

It is worth noting that for a fixed fiber transmission distance, more than one order magnitude BER improvement is achieved, while the CNN or BCNN scheme is only trained once over the entire measured optical power range. Therefore, the training process, which generally requires high computation cost, is significantly reduced with our proposed schemes. Specifically, compared to the FC-NN, the number of training iterations of the proposed CNN and BCNN decision schemes is reduced by about 50 \%, as shown by Fig.\ref{fig:training_result}. The required size of the training dataset is also significantly reduced in the proposed CNN and BCNN schemes, as shown by the BER performance with respect to the size of training dataset in Fig.\ref{fig:ber_vs_datasize}.
It is clear that when the size of training dataset increases, the BER performance improves before becoming stable, which represents the learning of almost all information carried by the training dataset. It can be seen from the figure that the size of training dataset required to reach stable BER is about 80,000 for the CNN and BCNN schemes, whilst it is about 120,000 for the FC-NN scheme. Hence, the required size of training dataset is reduced by over 30 \% in the CNN and BCNN schemes proposed here. Therefore, the computation costs of the proposed CNN and BCNN based decision schemes have been reduced significantly compared to previously studied FC-NN scheme.

One critical problem that machine learning algorithms typically encounter is overfitting. The overfitting issue is overcome in this paper through three major ways. Firstly, batch normalization technique was implemented, which has been shown to be highly effective in avoiding overfitting \cite{BN}; secondly, truly random data was used instead of PRBS data, which avoids the learning of data pattern; and thirdly, the feature sizes of input layer in the proposed neural networks were relatively small.

\section{Conclusion}
In this paper, we have proposed and investigated CNN and BCNN based decision schemes in 60 GHz mm-wave RoF system to compensate both linear and nonlinear impairments caused by signal modulation, transmission and detection, to improve the BER performance. Results have shown that by implementing CNN and BCNN at the receiver end, significantly improved BER performance has been achieved, and the improvement has been larger for longer distances. The performance of proposed CNN and BCNN decision schemes has also been compared with the previously studied FC-NN decision scheme. In addition to the improved BER performance, the computation cost of the proposed CNN and BCNN is significantly reduced, where the number of training iteration is reduced by about 50 \% and the size of training dataset is reduced by over 30 \%. In addition, the BER improvement enabled by the proposed CNN and BCNN has been achieved with only one training process over the entire measured optical power range (over 3.5 dB). Therefore, the computation cost of implementing neural networks decision schemes in mm-wave RoF systems is further reduced substantially.

\bibliographystyle{unsrt}  
\bibliography{references}  






\end{document}